\documentclass[a4paper,amsmath,amssymb,aps,prb,twocolumn,superscriptaddress]{revtex4-1}

\usepackage{graphicx}
\usepackage{dcolumn}
\usepackage{bm}
\usepackage{array}
\usepackage{xcolor}
\usepackage{longtable}
\usepackage{subfigure}

\usepackage[utf8]{inputenc}
\usepackage[T1]{fontenc}

\begin{document}

\title{Correlation strength, orbital-selective incoherence, and local moments formation in the magnetic MAX-phase Mn$_2$GaC}

\author{H. J. M. J\"onsson}
\thanks{Current Address: Department of Physics and Astronomy, Uppsala University, Uppsala, Sweden}
\affiliation{Theoretical Physics, Department of Physics, Chemistry and Biology (IFM), Link\"oping University, SE-581 83, Link\"oping, Sweden}

\author{M. Ekholm}
\affiliation{Theoretical Physics, Department of Physics, Chemistry and Biology (IFM), Link\"oping University, SE-581 83, Link\"oping, Sweden}

\author{I. Leonov}
\affiliation{M.N. Miheev Institute of Metal Physics, Russian Academy of
Sciences, 620108 Yekaterinburg, Russia}
\affiliation{Ural Federal University, 620002 Yekaterinburg, Russia}
\affiliation{Materials Modeling and Development Laboratory, National University of Science and Technology ‘MISiS’, Moscow, 119049, Russia}

\author{M. Dahlqvist}
\affiliation{Materials Design, Department of Physics, Chemistry and Biology (IFM), Link\"oping University, SE-581 83, Link\"oping, Sweden}

\author{J. Rosen}
\affiliation{Materials Design, Department of Physics, Chemistry and Biology (IFM), Link\"oping University, SE-581 83, Link\"oping, Sweden}

\author{I. A. Abrikosov}
\affiliation{Theoretical Physics, Department of Physics, Chemistry and Biology (IFM), Link\"oping University, SE-581 83, Link\"oping, Sweden}
\affiliation{Materials Modeling and Development Laboratory, National University of Science and Technology ‘MISiS’, Moscow, 119049, Russia}

\begin{abstract}
We perform a theoretical study of the electronic structure and magnetic properties of the prototypical magnetic MAX-phase Mn$_2$GaC with the main focus given to the origin of magnetic interactions in this system. 
Using the density functional theory+dynamical mean-field theory (DFT+DMFT) method we explore the effects of electron- electron interactions and magnetic correlations on the electronic properties, magnetic state, and spectral weight coherence of paramagnetic and magnetically-ordered phases of Mn$_2$GaC. We also benchmark the DFT-based disordered local moment approach for this system by comparing the obtained electronic and magnetic properties with that of the DFT+DMFT method.
%
Our results reveal a complex magnetic behavior characterized by a near degeneracy of the ferro- and antiferromagnetic configurations of Mn$_2$GaC, implying a high sensitivity of its magnetic state to fine details of the crystal structure and unit-cell volume, consistent with experimental observations. 
We observe robust local-moment behavior and orbital-selective incoherence of the spectral properties of Mn$_2$GaC, implying the importance of orbital-dependent localization of the Mn $3d$ states.
We find that Mn$_2$GaC can be described in terms of local magnetic moments, which may be modeled by DFT with disordered local moments. However, the magnetic properties are dictated by the proximity to the regime of formation of local magnetic moments, in which the localization is in fact driven by the Hund’s exchange interaction, and not the Coulomb interaction.
\end{abstract}

\maketitle

\section{Introduction}

MAX-phases are a promising class of functional materials with generic chemical formula M$_{n+1}$AX$_n$ and hexagonal crystal structure which consists of layers of C or N ('X') and transition metal ('M') atoms, interconnected by the layers of the A-group atoms.
First discovered in the 1960s \cite{jrn:rohde60,jrn:nowotny71} and later rediscovered in
1996 \cite{jrn:barsoum96} these materials possess intriguing physical behavior combining the properties typical for ceramics, such as high hardness, and those of metallic systems, e.g., good electrical and thermal conductivity\cite{jrn:eklund10}. This combination of the physical and mechanical properties, as well as being easily machined,\cite{jrn:barsoum11} makes MAX phases promising for numerous applications, such as thin film coatings for
low friction surfaces, electrical contacts, and heat
exchangers \cite{jrn:eklund10,jrn:barsoum11}.

\begin{figure}
\includegraphics[width=0.35\textwidth]{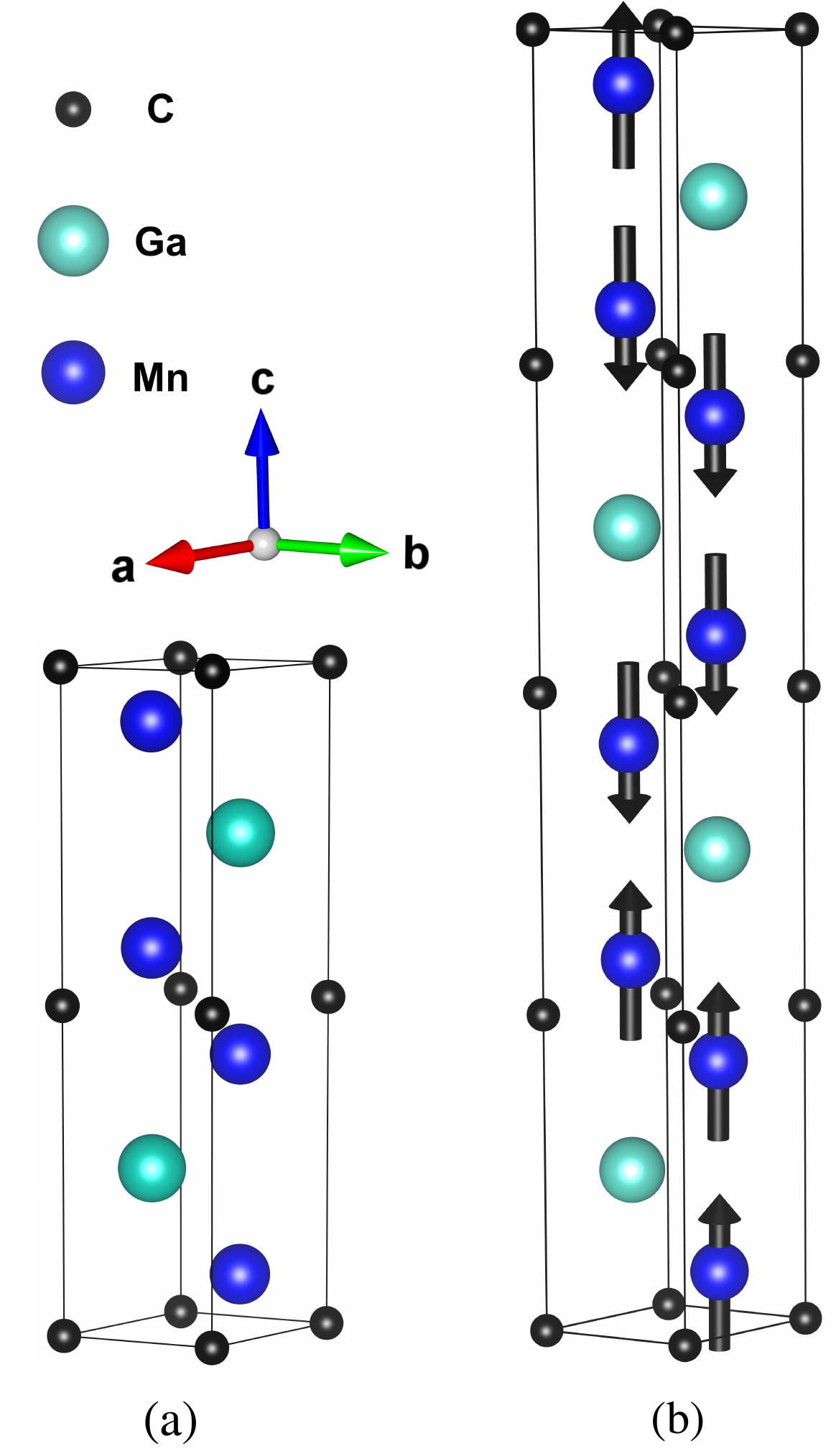}
\caption{(a) Crystal structure of Mn$_2$GaC. 
The space group is $P6_3/mmc$ with atomic positions: Mn at $4f$ $(\frac{1}{3},\frac{2}{3},0.0821)$, Ga at $2d$ $(\frac{1}{3},\frac{2}{3},\frac{3}{4})$, and C at $2a$ $(0,0,0)$.
(b) An AFM configuration of Mn$_2$GaC, showing the relative orientation of magnetic moments.
\label{fig:structure}}
\end{figure}

In 2013, the first experimental realization of a magnetic MAX-phase of Mn-doped Cr$_2$GeC was presented, synthesized as a heteroepitaxial single crystal thin film with excellent structural quality \cite{jrn:ingason13,jrn:lin13}.
In 2014, the long-range magnetically ordered Mn$_2$GaC MAX-phase was predicted by \emph{ab initio} band structure methods and then subsequently
synthesized \cite{jrn:ingason14}. It was the first magnetic MAX-phase with only
Mn atoms as the M-element. 
Its crystal structure is illustrated in Fig.\ \ref{fig:structure}.
The layered structure of MAX-phases leads to
a variety of possible magnetic structures that offers an intriguing perspective for applications of MAX-phases in spintronics and magnetic refrigeration \cite{jrn:ingason16iop}.

\emph{Ab initio} band-structure calculations, based on density functional theory \cite{jrn:hohenbergkohn} (DFT) within the local density approximation \cite{jrn:kohnsham,jrn:perdew92} (LDA) or generalized gradient approximation with the Perdew–Burke–Ernzerhof functional (PBE), \cite{jrn:perdew96} propose the existence of robust local magnetic moments in Mn$_2$GaC at low
temperatures,\cite{jrn:thore16,jrn:dahlqvist16} with a ferromagnetic (FM) ground state.
In contrast to this, neutron reflectometry in combination with DFT calculations suggests the formation of long-range antiferromagnetic (AFM) order in epitaxial thin films of Mn$_2$GaC \cite{jrn:ingason16}.
Moreover, the magnetic order was found to be very sensitive to small changes of the lattice
volume \cite{jrn:dahlqvist16}. Upon a slight expansion of the lattice, it leads to a transition from the FM to AFM state with a canted spin structure forming spin spirals with zero net magnetization. The theoretical calculations were consistent with the experimentally observed high sensitivity of the magnetic state of Mn$_2$GaC on temperature around room temperature \cite{jrn:dahlqvist16}.

The complicated behavior of the ordered magnetic structure of Mn$_2$GaC has also been observed in experiments\cite{jrn:ingason16,jrn:novoselova18}. 
In particular, above $\sim$507 K there appears to be a transition from a collinear AFM to a paramagnetic (PM) state, while below 214 K it makes a transition from collinear AFM to a canted AFM state.
In contrast to previously suggested local moment behavior, experimental studies of the MAX-phase (Cr,Mn)$_2$GaC with up to 25\% Mn report the FM state due to itinerant electrons, ruling out localized magnetic moments residing on the Mn sites \cite{jrn:liu14}. This raises a question about understanding of magnetic moments in Mn$_2$GaC, in terms of itinerant or localized electronic states, which poses a challenge for an accurate microscopic description of the electronic and magnetic properties of Mn$_2$GaC.

In this work, we explore the electronic structure and magnetic properties of Mn$_2$GaC using DFT+dynamical mean-field theory (DFT+DMFT) approach for strongly correlated systems \cite{jrn:georges96,RevModPhys.78.865,jrn:aichhorn09}.
By using the DFT+DMFT approach it becomes possible to proceed beyond a static mean-field treatment of the electron-electron interactions in DFT, to capture correlated electron phenomena such as a quasiparticle behavior, orbital-selective band
mass renormalizations, coherence-incoherence
crossover of the spectral weight, to consider temperature-induced (local) quantum fluctuations, and to explain the Mott transition \cite{book:martin16,jrn:georges96,RevModPhys.78.865,PhysRevB.81.045117,PhysRevLett.106.106405,PhysRevB.86.035152,PhysRevB.87.125138,Pourovskii_2017,PhysRevB.92.085142,PhysRevB.94.155135,PhysRevB.101.245144,PhysRevX.8.031059,PhysRevLett.121.197001,PhysRevLett.118.167003,PhysRevLett.115.106402}.
Our results suggest that Mn$_2$GaC is a correlated metal sitting near to the regime of formation of local magnetic moments, in which localization is driven by the on-site Hund's exchange coupling $J$ (in contrast to the Hubbard repulsion $U$ value).

\section{Computational details}\label{sec:methods}
We employ the state-of-the-art fully self-consistent in charge density DFT+DMFT method \cite{RevModPhys.78.865,jrn:georges96} to examine the electronic structure and magnetic properties of paramagnetic and magnetically-ordered states of Mn$_2$GaC.
In our DFT+DMFT calculations we use two independent DFT+DMFT implementations. The first is provided in the Toolbox for research in interacting quantum systems (TRIQS) package for DMFT, which is interfaced with the all-electron Wien2k DFT code \cite{jrn:triqs,book:wien2k,blaha20,jrn:aichhorn09,jrn:aichhorn11,jrn:dfttools3}.
We used LDA and set $R_{\mathrm{MT}}K_{\mathrm{max}}=8.0$. In DFT+DMFT calculations we consider the Mn $3d$ valence states as correlated orbitals by constructing a basis set of atomic-centered Wannier functions within the energy window $[-7.8, 6.4]$ eV (Fermi energy $E_F=0$). \cite{jrn:anisimov05,JPhysCondMat.20.135227,jrn:aichhorn11} We use the continuous-time hybridization expansion quantum Monte Carlo algorithm within the full rotationally-invariant form in order to solve the realistic many-body problem in DMFT,\cite{jrn:ctqmc1,jrn:ctqmc2,phd:ctqmc3,jrn:ctqmc4,phd:ctqmc5,RevModPhys.83.349} 
as implemented in the CTHYB package. \cite{jrn:cthyb}

In the second case, we use DFT+DMFT implemented within a plane-wave pseudopotential formalism \cite{PhysRevB.92.085142,PhysRevB.94.155135,PhysRevX.8.031059,PhysRevB.101.245144} in DFT \cite{Giannozzi_2009}, combined with the continuous-time hybridization expansion (segment) quantum Monte Carlo algorithm in DMFT \cite{RevModPhys.83.349,RevModPhys.78.865,jrn:georges96}. In this approach the Coulomb interaction is treated in the density-density approximation, neglecting by spin-flip and pair-hopping terms in the multiorbital Hubbard Hamiltonian. We use the generalized gradient approximation with the PBE functional in DFT \cite{jrn:perdew96}. In our DFT+DMFT calculations we explicitly include the Mn $3d$, Ga $4s$ and $4p$, and C $2p$ valence states, by constructing a basis set of atomic-centered Wannier functions within the energy window spanned by these bands \cite{jrn:anisimov05,JPhysCondMat.20.135227}. This allows us to take into account charge transfer between the partially occupied Mn $3d$, Ga $4s$ and $4p$, and C $2p$ valence states, accompanied by the strong on-site Coulomb correlations of the Mn $3d$ electrons. The DFT+DMFT calculations are performed with full self-consistency over the charge density.

Using both DFT+DMFT approaches we compute the electronic structure, magnetic properties, quasiparticle mass renormalizations $m^*/m$, and spin-spin correlation function $\chi(\tau)$ of the PM, FM and AFM states of Mn$_2$GaC as shown in Fig.~\ref{fig:structure}. We note that in our spin-polarized DFT+DMFT calculations the nonmagnetic DFT was employed (exchange splitting due to magnetism appears in DMFT). We take $U=3.8$ eV for the average Hubbard interaction, as estimated previously using constrained random phase approximation \cite{jrn:miyake08}, and the Hund's exchange coupling $J=0.95$ eV. In addition, we explore the effects of correlation strength on the electronic and magnetic properties of Mn$_2$GaC taking the different values of the Coulomb repulsion and Hund's exchange couplings $U=5.3$ and 6.9 eV, and $J=0.5$ eV, respectively.\cite{jrn:kunes08,tomczak10,vaugier12,petocchi20}

In order to analyse the degree of localization of the Mn $3d$ electrons of the PM, FM, and AFM Mn$_2$GaC we compute the local spin-spin correlation function $\chi(\tau)= \langle \hat{m}_z(\tau)\hat{m}_z(0)\rangle$ within DMFT, where $\hat{m}_z(\tau)$ is the instantaneous magnetization on the Mn $3d$ site at the imaginary time $\tau$ (the latter denotes an imaginary-time evolution ranging from 0 to $\beta=1/k_\mathrm{B}T$ in the path integral formalism) \cite{jrn:georges96,RevModPhys.78.865}.
In both DFT+DMFT calculations, the fully localized double-counting correction evaluated from the self-consistently determined local occupations was used. The spin-orbit coupling is neglected in our calculations.
In order to obtain the self-energy on the real axis we employ the maximum entropy method, as implemented in the MaxEnt package \cite{kraberger17} and Pad\'e approximants.

In Fig.\ \ref{fig:structure}(a) we display the non-magnetic two-formula-units cell (contains eight atoms) of Mn$_2$GaC with the lattice volume of 40.87 \AA$^3$/f.u. and $c/a=4.3$, used in the DFT+DMFT calculations
(the corresponding lattice parameters are evaluated from optimization of the unit-cell shape within nonmagnetic DFT).
In addition, we compute the electronic and magnetic properties of PM Mn$_2$GaC with DFT-PBE for magnetically disordered supercells of Mn$_2$GaC using the projector-augmented waves technique,\cite{jrn:blochl94} as implemented in the Vienna \emph{ab-initio} simulation package (VASP) \cite{jrn:vasp1,jrn:vasp2}.
To this end, we set up a 128-atom supercell consisting of $4\times 4\times 1$ unit cells.
In order to model the PM state we distributed the up/down Mn magnetic moments according to the special quasirandom structure (SQS) technique. \cite{jrn:zunger90,jrn:alling10}
We minimize the short-range order parameter\cite{cowley50}  
$ \alpha_i  = 1 - \frac{P_i(\uparrow | \downarrow)}{c^\downarrow}$, 
where $P_i(\uparrow | \downarrow)$ is the average conditional probability of finding a spin-down Mn-atom in the $i$-th coordination shell of a spin-up atom on the Mn sublattice; $c^\downarrow=0.5$ is the concentration of spin-down atoms on the sublattice.
After self-consistency, one of the 64 Mn magnetic moments had flipped.
Nevertheless, the absolute value of the short-range order parameters was below 0.08 for the first eight nearest-neighbor shells.
In these calculations we used a $5\times 5 \times 5$ $\mathbf{k}$-point mesh and the plane wave cutoff 400 eV.
The unit-cell volume was taken to be $44.66$ \AA$^3$/f.u.\ and $c/a=4.29$ as obtained from structural optimization of Mn$_2$GaC within static DFT-SQS (for details see Ref.~\onlinecite{jrn:dahlqvist16}).

\section{Results}
\label{sec:res}

\subsection{FM and AFM long-range ordered magnetic states}

\begin{figure}
    \centering
    \includegraphics[width=0.5\textwidth]{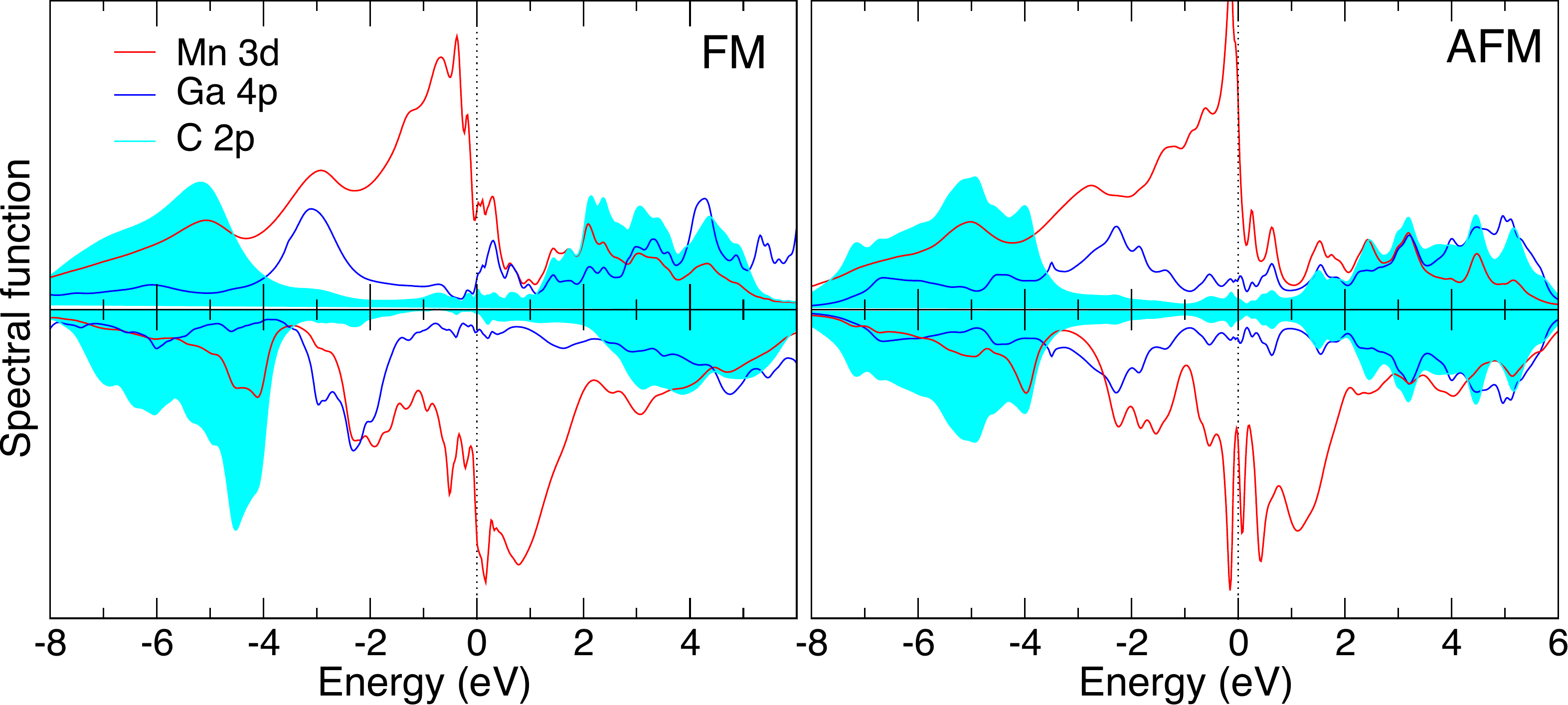}
    \caption{Mn $3d$, Ga $4p$, and C $2p$ spectral functions of the FM and AFM phases of Mn$_2$GaC obtained within the spin-polarized DFT+DMFT calculations at $T=290$ K.}
    \label{fig:spectra}
\end{figure}

\begin{figure}
    \centering
    \includegraphics[width=0.45\textwidth]{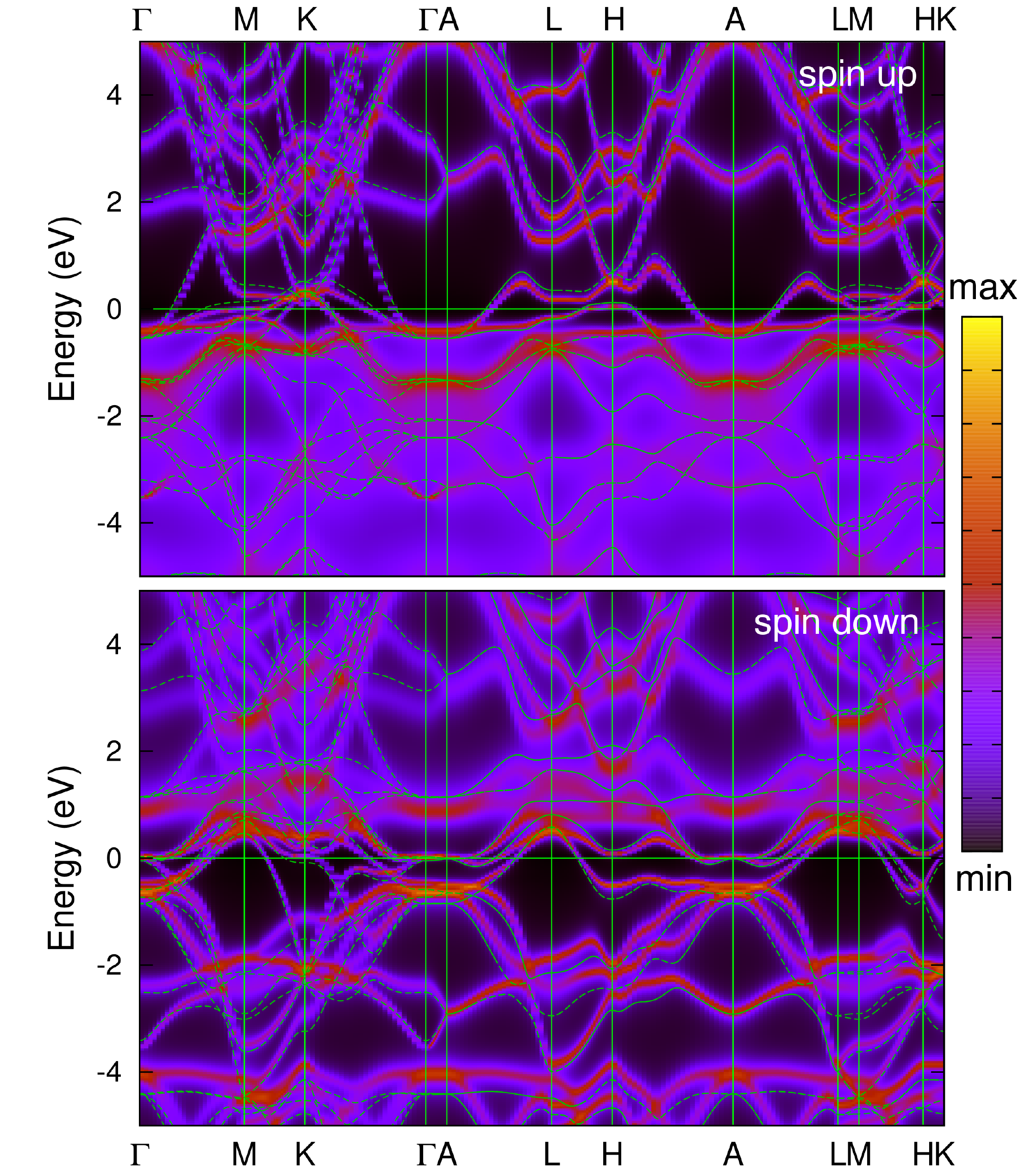}
    \caption{$\mathbf{k}$-resolved spectral function of FM Mn$_2$GaC calculated by DFT+DMFT at $T=290$ K in comparison to the spin-polarized DFT results (shown in green).}
    \label{fig:FM_arpes}
\end{figure}

\begin{figure}
\includegraphics[width=0.45\textwidth]{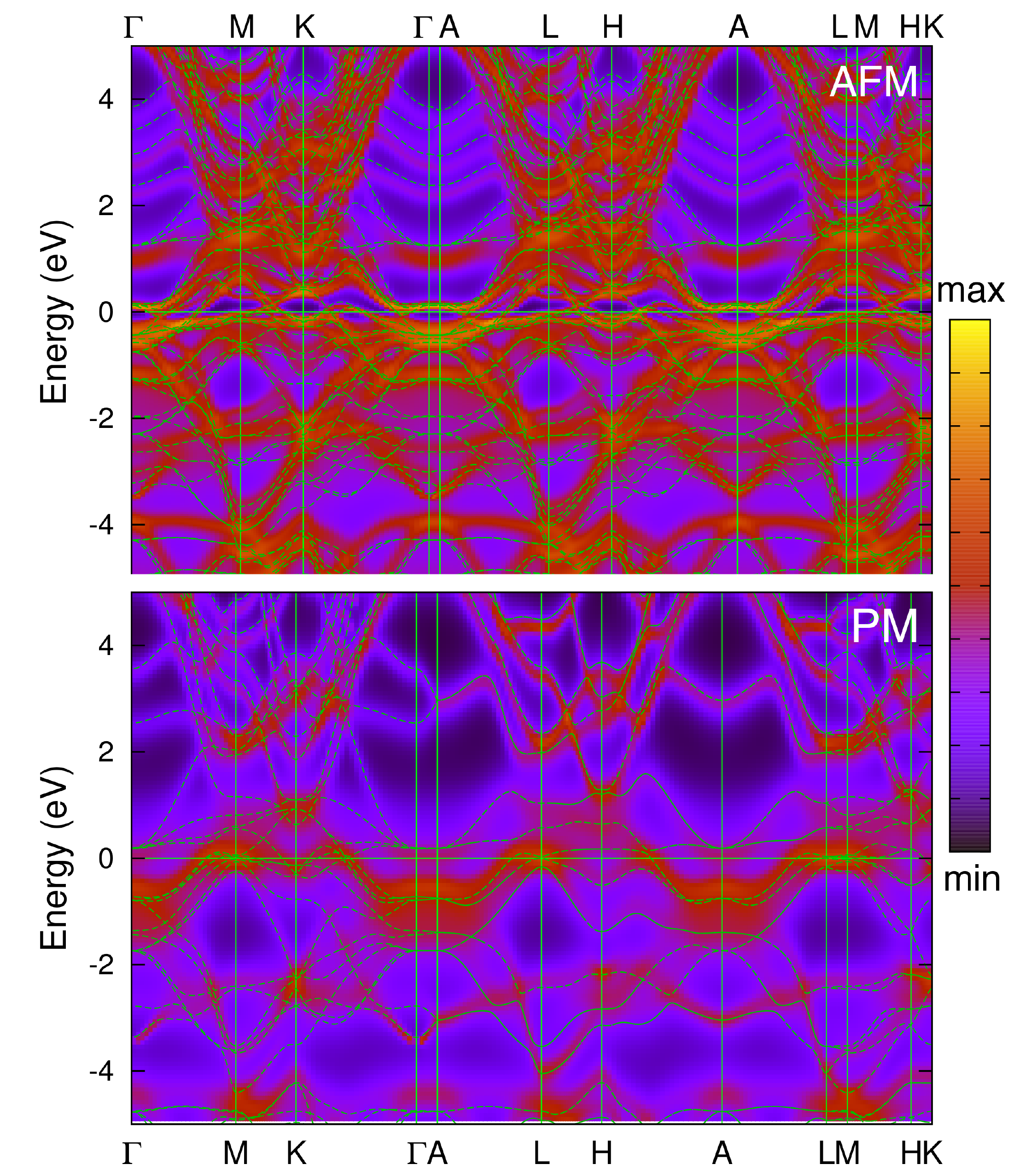} 
\caption{
$\mathbf{k}$-resolved spectral function of the AFM and PM phases of Mn$_2$GaC calculated by DFT+DMFT in comparison to the ``bare'' Kohn-Sham band structure obtained within DFT (shown in green).
\label{fig:spectrum}}
\end{figure}

We start by computing the electronic structure and magnetic properties of the FM and AFM phases of Mn$_2$GaC at a temperature $T=193$~K using the spin-polarized DFT+DMFT method with $U=3.8$ eV and $J=0.95$ eV. We note that both DFT+DMFT schemes discussed above give nearly identical results. 
In Fig.~\ref{fig:spectra} we display our DFT+DMFT results for the Mn $3d$, Ga $4p$, and C $2p$ spectral functions of Mn$_2$GaC. Our results for the {\bf k}-resolved spectral functions $A(\mathbf{k},\omega)=-\frac{1}{\pi}\mathrm{Tr}G(\mathbf{k},\omega)$ of the FM and AFM phases of Mn$_2$GaC are shown in Figs.~\ref{fig:FM_arpes} and \ref{fig:spectrum}.

We find a metallic solution with a long-range magnetic ordering of the Mn ions.
The Mn $3d$ states are strongly hybridized with the C $2p$ and Ga $4p$ states and form a broad band of about 8 eV bandwidth located at the Fermi level. The occupied C $2p$ and Ga $4p$ states appear in between about -8 to -4 eV and near to -4 eV below $E_F$, respectively. The unoccupied C $2p$ and Ga $4p$ bands sit above 2 eV. The spectral functions of FM and AFM Mn$_2$GaC show a pronounced $\sim$1--2 eV splitting of the Mn $3d$ spin-up and spin-down states due to the magnetic exchange (see Fig.~\ref{fig:spectra}).

Our results for the self-energy show a typical Fermi-liquid-like behavior, though with a notable damping of quasiparticles of the Mn $t_2$ ($a_{1g}$ and $e_g^\pi$) orbitals in the AFM state ($\mathrm{Im}[\Sigma(0^+)]$$\sim$0.01--0.03) eV). The latter is seen as (orbital-selective) incoherence of the {\bf k}-resolved spectral weight of the Mn $3d$ states near the Fermi level in Mn$_2$GaC (see Figs.~\ref{fig:FM_arpes} and \ref{fig:spectrum}), implying the importance of electronic correlations \cite{PhysRevB.100.245109,PhysRevB.103.155115,PhysRevB.96.195121,PhysRevLett.121.197001,PhysRevLett.118.167003,PhysRevLett.115.106402,PhysRevB.97.115165,PhysRevB.96.035137}.
Moreover, we evaluate the quasiparticle mass enhancement $m^*/m=1-\partial Im \Sigma(\omega)/\partial \omega|_{\omega=0}$ using extrapolation of the self-energy $\Sigma(\omega)$ to $\omega = 0$ eV, which gives a quantitative measure of the correlation strength. For the long-range FM and AFM ordered states we obtain a moderately correlated metal with $m^*/m \sim1.4$ and 1.6, respectively, with a weak orbital-dependence of $m^*/m$. In fact, the electronic band structure obtained by DFT+DMFT closely follows the spin-polarized DFT results, with a notable bandwidth renormalizaton and significant broadening (incoherence) of the spectral weight caused by correlation effects.

\begin{figure}
\includegraphics[angle=0,width=0.5\textwidth]{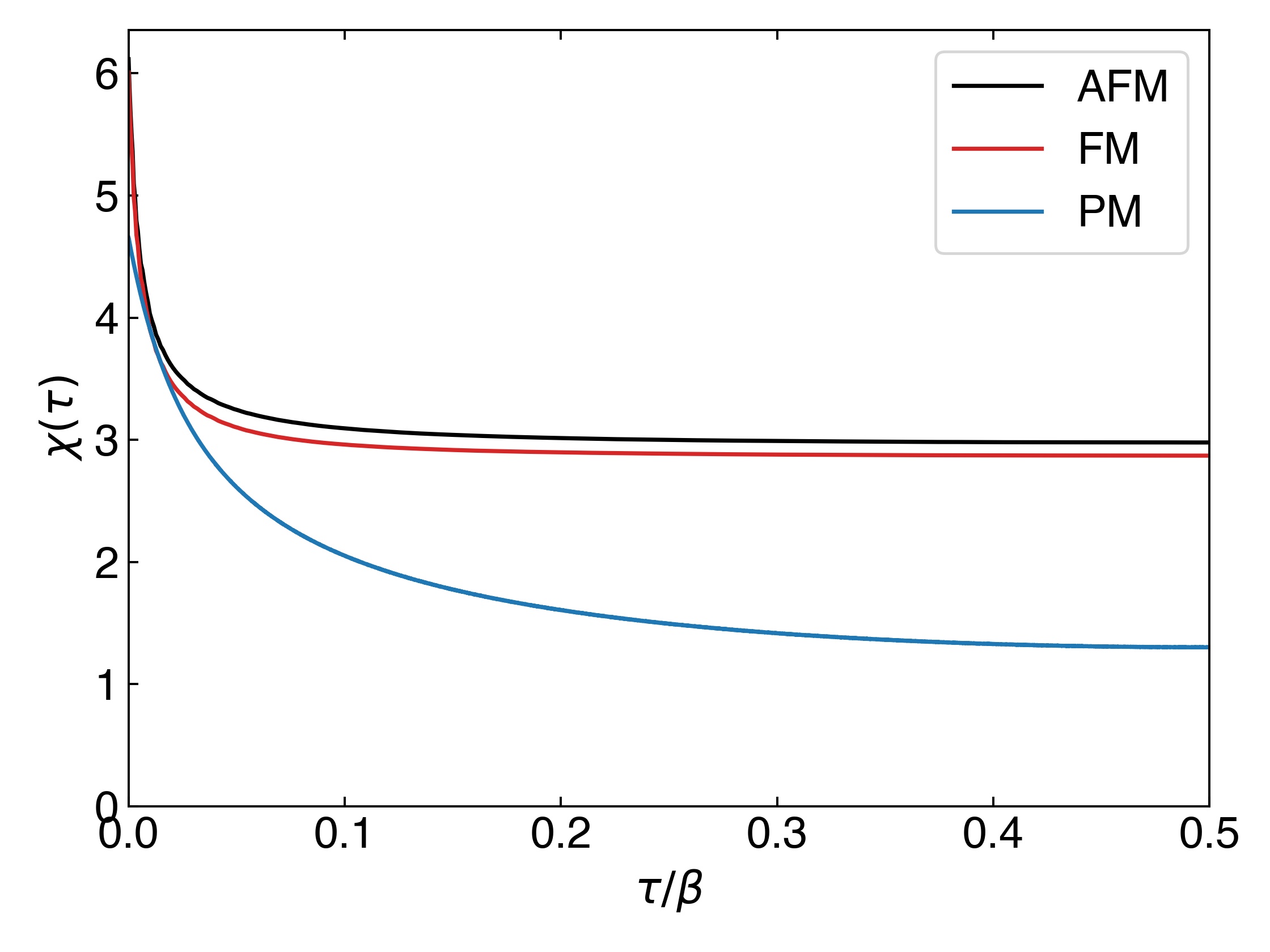}
\caption{Local spin-spin correlation function $\chi(\tau)= \langle \hat{m}_z(\tau)\hat{m}_z(0)\rangle$ calculated by DFT+DMFT (with $U=3.8$ eV and $J=0.95$ eV) as a function of the imaginary time $\tau$. Our DFT+DMFT results for the FM and AFM phases at $T\sim 193$ K ($\beta=60$ eV$^{-1}$) and the PM phase at $T\sim 1160$ K ($\beta=10$ eV$^{-1}$) are shown.}
\label{fig:spinautocorr}
\end{figure}

\begin{table}
\caption{\label{tab:mag} Long-range ordered $\langle \hat{m}_z \rangle$, fluctuating $M_\mathrm{loc}$, and instantaneous $\sqrt{\langle \hat{m}^2_z\rangle}$ magnetic moments of Mn ions of the FM, AFM, and PM phases of Mn$_2$GaC as obtained by DFT+DMFT at different temperatures. The Hubbard interaction $U=3.8$ eV. In the DFT+DMFT results at $T=116$ K the Hund's exchange is taken $J=0.5$ eV, while at $T=193$ K and 1160 K $J=0.95$ (see text).}
\begin{ruledtabular}
\begin{tabular}{cccccc}
   & \multicolumn{2}{c}{116 K} & \multicolumn{2}{c}{193 K} & 1160 K\\
 &  AFM & PM & FM & AFM & PM \\
\hline 
$\langle \hat{m}_z\rangle$ & 0.12 & 0 & 1.70 & 1.73 & 0 \\
$M_{\textrm{loc}} $  & 0.31 & 0.3 & 1.72 & 1.76 & 1.4  \\
$\sqrt{\langle \hat{m}_z^2\rangle}$ & 1.8 & 1.81 & 2.2 & 2.2 & 2.3 \\ 
\end{tabular}
\end{ruledtabular}
\end{table}

The calculated long-range ordered magnetic moments for the FM and AFM states (at 193 K) $\sim$1.70$\mu_B$ and 1.73 $\mu_B$ per Mn ion, respectively, are comparable to those found within spin-polarized DFT (PBE) for the theoretical equilibrium volume, $1.59\mu_B$ and $1.83\mu_B$ \cite{jrn:dahlqvist16} (see Table~\ref{tab:mag}). At the same time, the instantaneous magnetic moment $\sqrt{\langle \hat{m}_z^2\rangle}$ is higher, $\sim$2.5$\mu_B$, implying the presence of significant (local) spin fluctuations. In order to quantify the robustness of magnetic moments of the Mn $3d$ states with respect to quantum fluctuations, we compute the local spin-spin correlation function $\chi(\tau)= \langle \hat{m}_z(\tau)\hat{m}_z(0)\rangle$ within DMFT. Our DFT+DMFT results for $\chi(\tau)$ are summarized in Fig.~\ref{fig:spinautocorr}. Our analysis of the local spin susceptibility suggests the proximity of Mn $3d$ moments to localization. In particular, while $\chi(\tau)$ is seen to slowly decay with the imaginary time at small $\tau$ (mainly for $\tau/\beta<0.05$), it is significant and nearly constant for $\tau \simeq \beta/2$, $\sim$3.0$\mu_B^2$, 
which implies the robustness of local magnetic moments in Mn$_2$GaC. Moreover, the calculated fluctuating local moments $M_\mathrm{loc}=[T\int_0^{1/T}d\tau \langle \hat{m}_z(\tau)\hat{m}_z(0)\rangle]^{1/2}$ are $\sim$1.72$\mu_B$ and $1.76\mu_B$ for the FM and AFM states, respectively. The latter are compatible with the long-range magnetic moment of $\sim$1.70-1.73$\mu_B$ per Mn ion in FM and AFM Mn$_2$GaC. 

Our DFT+DMFT total-energy calculations (with $U=3.8$~eV and $J=0.95$~eV) suggest a near degeneracy of the FM and AFM states of Mn$_2$GaC. In particular, we find that at 193~K the AFM state is energetically favorable, with a small total energy difference between the FM and AFM state of $\sim$4~meV/f.u., whereas at 293~K it differs by $-5$~meV/f.u., with the FM state being most stable. This behavior suggests a high sensitivity of the magnetic state of Mn$_2$GaC, e.g., to fine details of its crystal structure, lattice volume, and temperature, in agreement with experimental observations. At the same time, the PM state is found to be highly energetically unfavorable with a total energy difference of $\sim$95 meV/f.u., suggesting strong magnetic exchange interactions in Mn$_2$GaC.

Moreover, we observe that the magnetic properties of Mn$_2$GaC depend very sensitively on the particular choice of the Hund's exchange coupling $J$. In fact, for a smaller value $J=0.5$~eV (and the same Hubbard interaction $U=3.8$~eV) both the FM and AFM states are found to be unstable and collapses to the PM state at room temperature. At $\sim$116~K for AFM Mn$_2$GaC we find long-range magnetic ordering with a weak static magnetic moment of $\sim$0.12$\mu_B$ per Mn ion. We note that at $\sim$116~K the FM phase is unstable and collapse in the PM state. The calculated instantaneous local moments are significantly smaller, $\sim$1.8$\mu_B$, than those for $J=0.95$~eV. In addition to this, our DFT+DMFT calculations with $U=3.8$~eV and $J=0.5$~eV of the local spin susceptibility $\chi(\tau)$ give a typical itinerant moment behavior, with $\chi(\tau)$ quickly decaying to zero with the imaginary time $\tau$. Our result for the fluctuating moment is $\sim$0.3$\mu_B$ (at $\sim$116~K).

\subsection{Paramagnetic phase}

\begin{figure}
    \centering
    \includegraphics[width=0.5\textwidth]{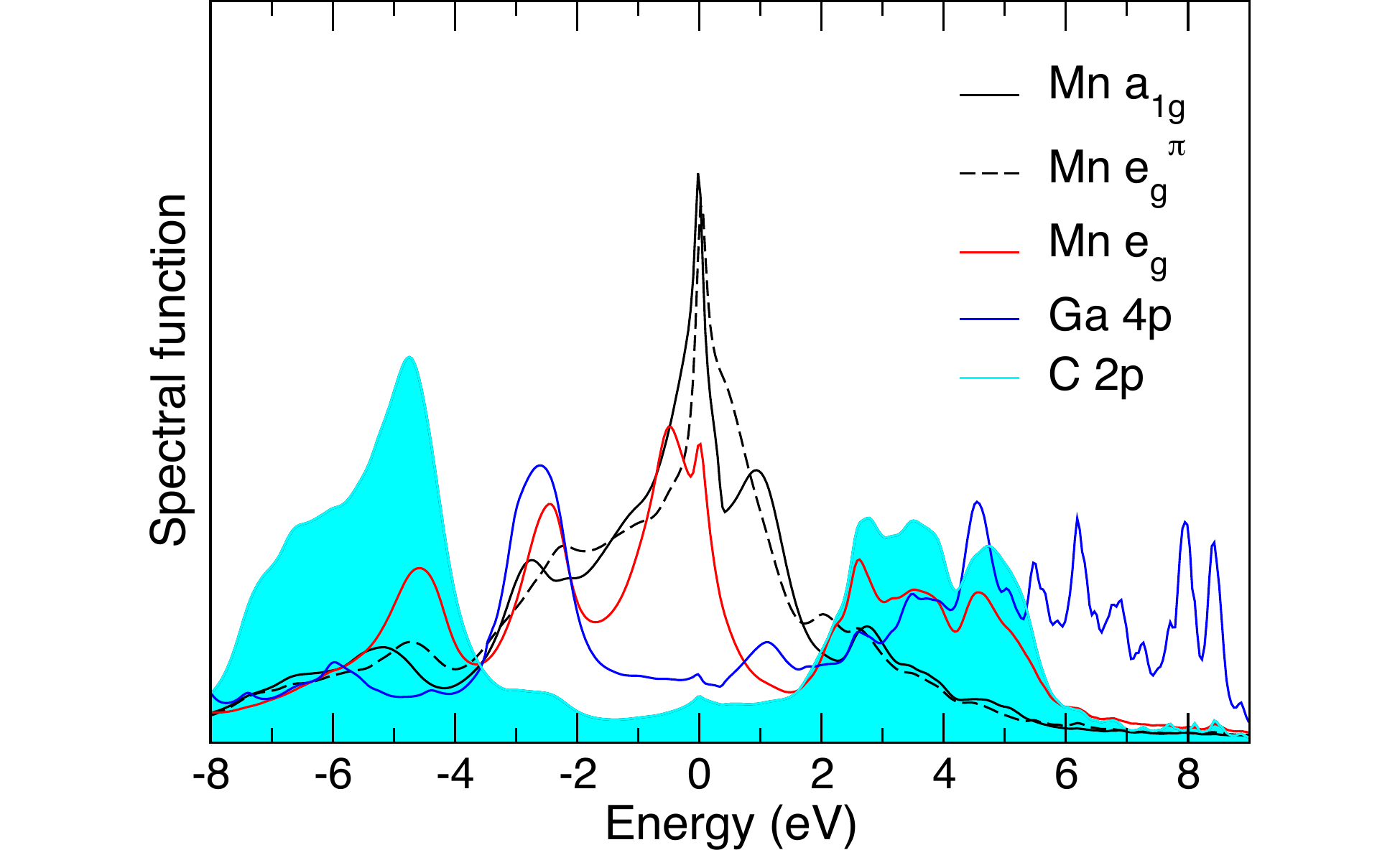}
    \caption{Orbitally-resolved Mn $3d$, Ga $4p$, and C $2p$ spectral functions of PM Mn$_2$GaC obtained by DFT+DMFT.}
    \label{fig:PM_spectra}
\end{figure}

Next, we compute the electronic structure and magnetic state of the PM phase of Mn$_2$GaC at a high electronic temperature $T \sim 1160$~K (well above the experimental magnetic ordering temperature of $\sim$507~K) using the DFT+DMFT method with $U=3.8$ eV and $J=0.95$ eV. In Fig.~\ref{fig:spectrum} (bottom panel) we display our results for the {\bf k}-resolved spectral function of PM Mn$_2$GaC in comparison with the ``bare'' Kohn-Sham band structure, calculated within nonmagnetic DFT. 
Our results for the orbitally-resolved Mn $3d$, Ga $4p$, and C $2p$ spectral functions of PM Mn$_2$GaC are shown in Fig.~\ref{fig:PM_spectra}.

\begin{figure}
    \centering
    \includegraphics[width=0.5\textwidth]{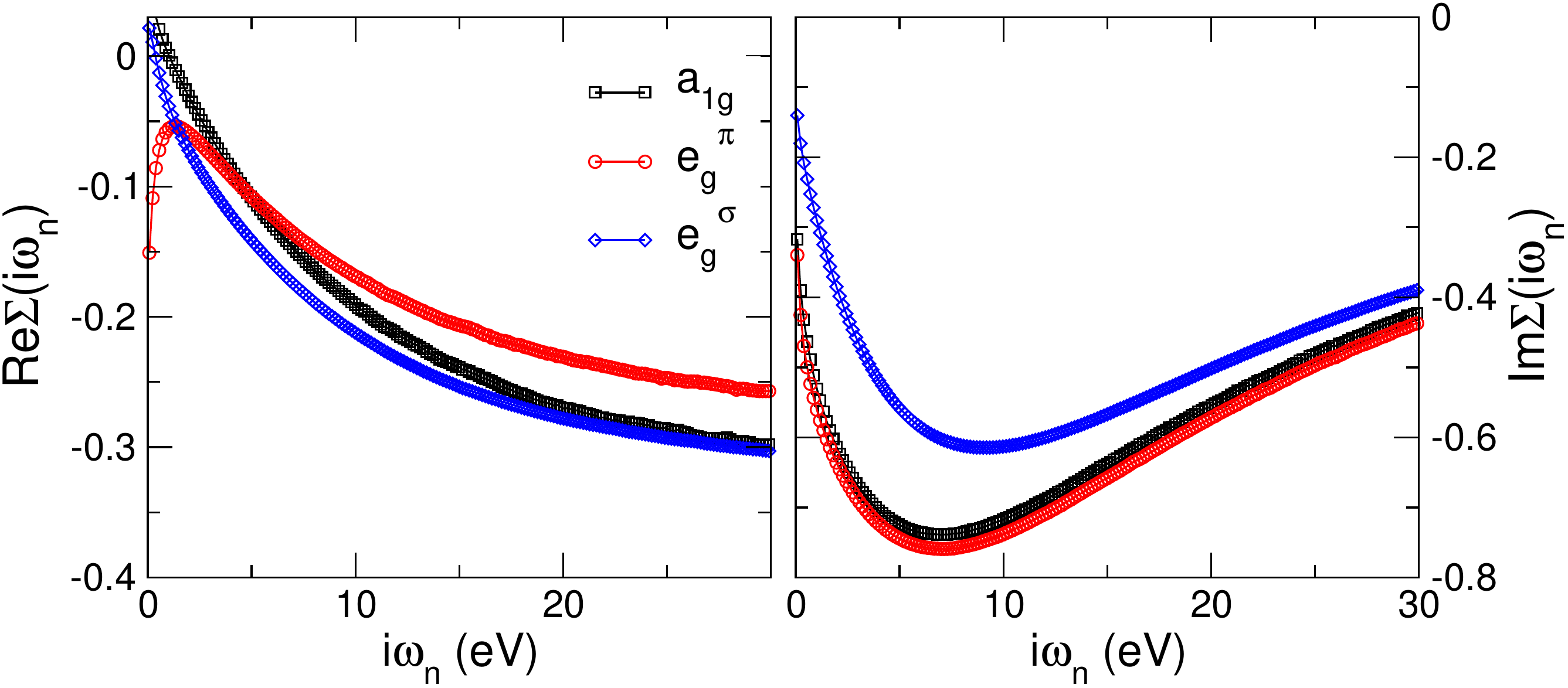}
    \caption{Orbitally-resolved Mn $3d$ self-energy on Matsubara contour $\Sigma(i\omega_n)$ of PM Mn$_2$GaC evaluated by DFT+DMFT.}
    \label{fig:PM_sigma}
\end{figure}

We obtain a strongly correlated metal with a pronounced quasiparticle peak at the Fermi level due to the Mn $a_{1g}$ and $e_g^\pi$ states. The Mn $e_g^\sigma$ states are seen to strongly hybridize with the Ga $4p$ and C $2p$ states and are of about 10 eV bandwidth. In contrast, the Mn $t_2$ bandwidth is more narrow of $\sim$6 eV, suggesting the importance of orbital-selective correlations. Mn$_2$GaC is characterized by a Fermi-liquid-like behavior of the self-energy and large damping (finite lifetime) of quasiparticles originating from the Mn $t_2$ states, $\mathrm{Im}[\Sigma(0^+)]\sim0.34$ eV in the PM state at 290~K, respectively (see Fig.~\ref{fig:PM_sigma}). 
For the Mn $e_g^\sigma$ states it is weaker, of $\sim$0.14 eV.
The latter is consistent with significant (orbital-dependent) incoherence of the spectral weight of the Mn $3d$ states near the Fermi level [see Fig.~\ref{fig:spectrum} (bottom panel)], implying the importance of electronic correlations \cite{PhysRevB.100.245109,PhysRevB.103.155115,PhysRevB.96.195121,PhysRevLett.121.197001,PhysRevLett.118.167003,PhysRevLett.115.106402,PhysRevB.97.115165,PhysRevB.96.035137}. This behavior persists even at relatively low temperature of $\sim$193~K (in the PM state). 
Moreover, our analysis of the orbitally resolved quasiparticle mass renormalizations yields $m^*/m \sim 2.6$ and 1.6 for the Mn $t_2$ ($a_{1g}$ and $e_g^\pi$) and $e_g^\sigma$ states, respectively, implying orbital selectivity of correlation effects in Mn$_2$GaC. The latter agrees well with sufficiently different bandwidth of the Mn $t_2$ and $e_g^\sigma$ states, as well as with orbital-selective incoherence of the Mn $3d$ states.
Our results for the instantaneous magnetic moment of the Mn ions is $\sim$2.3$\mu_B$. This value is larger than the long-range ordered magnetic moment obtained by DFT+DMFT for the FM and AFM phases of Mn$_2$GaC at $T \sim 193$~K, $\sim$1.7~$\mu_B$ per Mn ion, due to temperature-induced quantum fluctuations.

In addition to this, our results for the local spin susceptibility $\chi(\tau)$ (with $U=3.8$ eV and $J=0.95$ eV) show the presence of local moment behavior in PM Mn$_2$GaC. $\chi(\tau)$ shows a slow decay with the imaginary time $\tau$ to a nearly flat region with $\chi \simeq 1.4\mu_B^2$ near $\tau \simeq \beta/2$, implying the robustness of local magnetic moments in the PM phase. Moreover, the calculated fluctuating moment $M_\mathrm{loc}$ is large, $\sim$1.4$\mu_B$, consistent with the regime of formation of local magnetic moments. Our results reveal a remarkable orbital-selective renormalization of the Mn $3d$ bands in PM Mn$_2$GaC, which suggests orbital-dependent localization of the Mn $3d$ states.
We notice a remarkable \emph{increase} of local moments (both instantaneous and fluctuating) in the low-temperature FM and AFM Mn$_2$GaC, suggesting an \emph{enhancement} of localization of the Mn $3d$ states in the FM and AFM phases of Mn$_2$GaC. On the other hand, this behavior is accompanied by a sizable decrease of the quasiparticle mass renormalizations, suggesting the importance of orbital-selective damping of quasiparticle coherence, i.e., an enhancement of the strength of electronic correlations.  

Interestingly, in contrast to a sharp dependence on the Hund's exchange $J$, we observe a rather weak change of the electronic structure and magnetic properties (e.g., local moments) upon a large variation of the Hubbard $U$ value from 3.8 to 5.3 and 6.9~eV. In particular, the instantaneous moments increase from $2.3\mu _B$ to $2.5\mu _B$, only by less than 9\% upon a change of the Hubbard $U$ from 3.8 to 6.9~eV (with $J=0.95$ eV). This anomalous dependence of the electronic and magnetic properties of Mn$_2$GaC on the Hund's coupling $J$, with a rather weak dependence on the Hubbard $U$ is reminiscent of that in Hund's metals \cite{PhysRevB.83.205112,PhysRevLett.107.256401,PhysRevB.100.085104,PhysRevLett.122.186401,AnnuRevCondensMatterPhys.4.137,AnnPhysics.405.265,jrn:yin11,jrn:deng19,jrn:Katanin21,jrn:haule09}. We conclude that the magnetic properties of Mn$_2$GaC are dictated by its proximity to the regime of formation of local magnetic moments, in which localization is driven by the Hund's exchange coupling $J$. This implies the crucial importance of (orbital-dependent) correlation effects for understanding the electronic and magnetic properties of Mn$_2$GaC.

\begin{figure}
    \centering
    \includegraphics[width=0.5\textwidth]{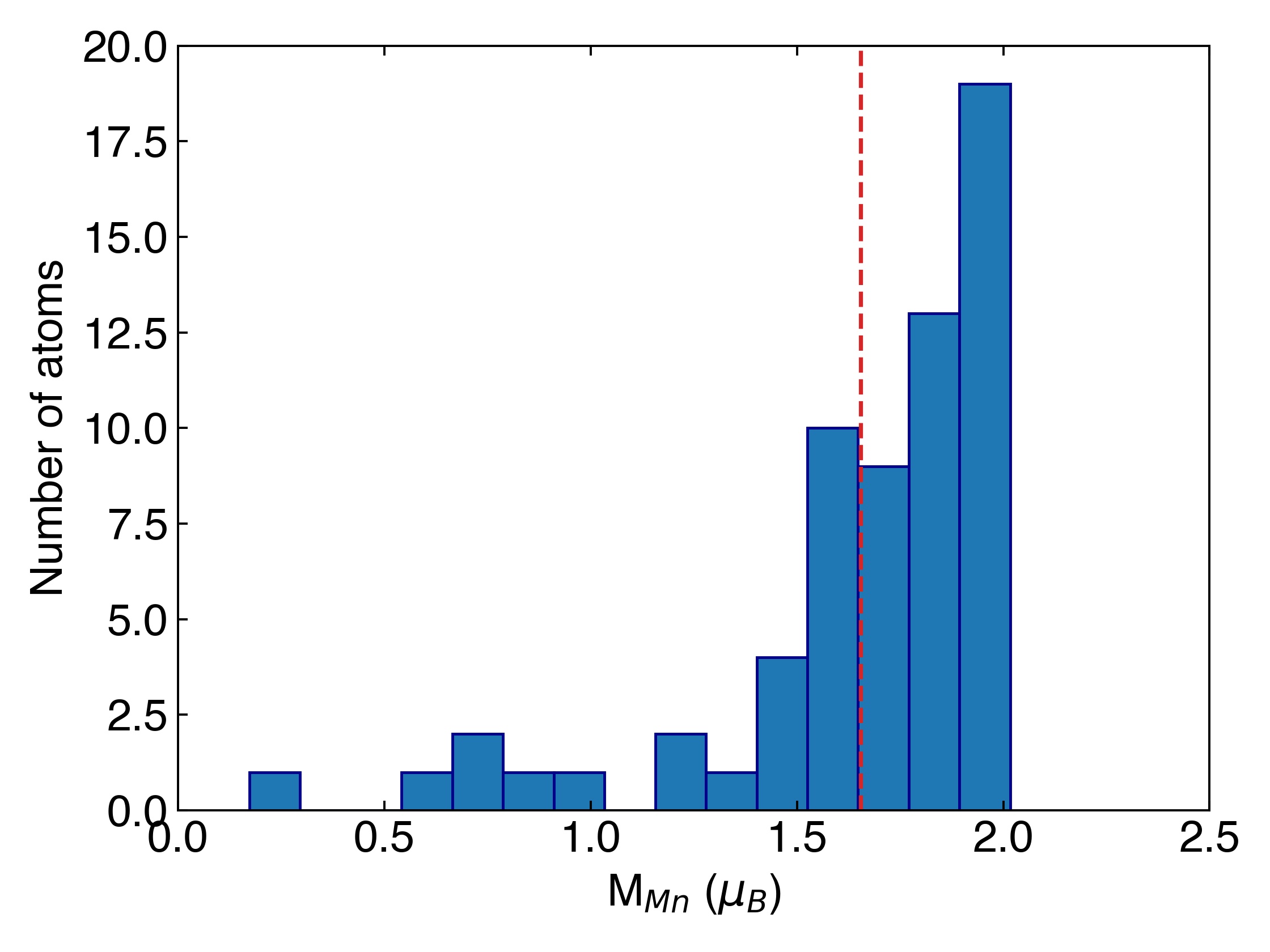}
    \caption{Distribution of Mn magnetic moments obtained from supercell DFT-DLM calculations with PBE.}
    \label{fig:momdist}
\end{figure}

To better establish the relation between our advanced DFT+DMFT calculations\cite{mandal19} with earlier simulations of Mn$_2$GaC carried out with DFT, we have performed DFT calculations of the magnetic properties of PM Mn$_2$GaC within the static disordered local moment (DLM) picture.
In the this approach, the quantum PM state is approximated by disordered local magnetic moments distributed over the Mn sites in the supercell of Mn$_2$GaC consisting of $4 \times 4 \times 1$ unit-cells to mimic complete magnetic disorder in the thermodynamic limit \cite{jrn:alling10}, although neglecting quantum fluctuations. In Fig.~\ref{fig:momdist} we display a histogram over the magnitudes of local magnetic moments of the Mn ions. 
It has a mean value of $\sim$1.7$\mu_B$, compatible to the fluctuating local magnetic moments obtained by DFT+DMFT, with a significant portion of magnetic moments deviating from the mean value.
This indicates a sensitivity of magnetic moments to the local environment, which is usually seen in itinerant electron magnets.\cite{jrn:james99}

\begin{figure}
\centering
\includegraphics[width=0.5\textwidth]{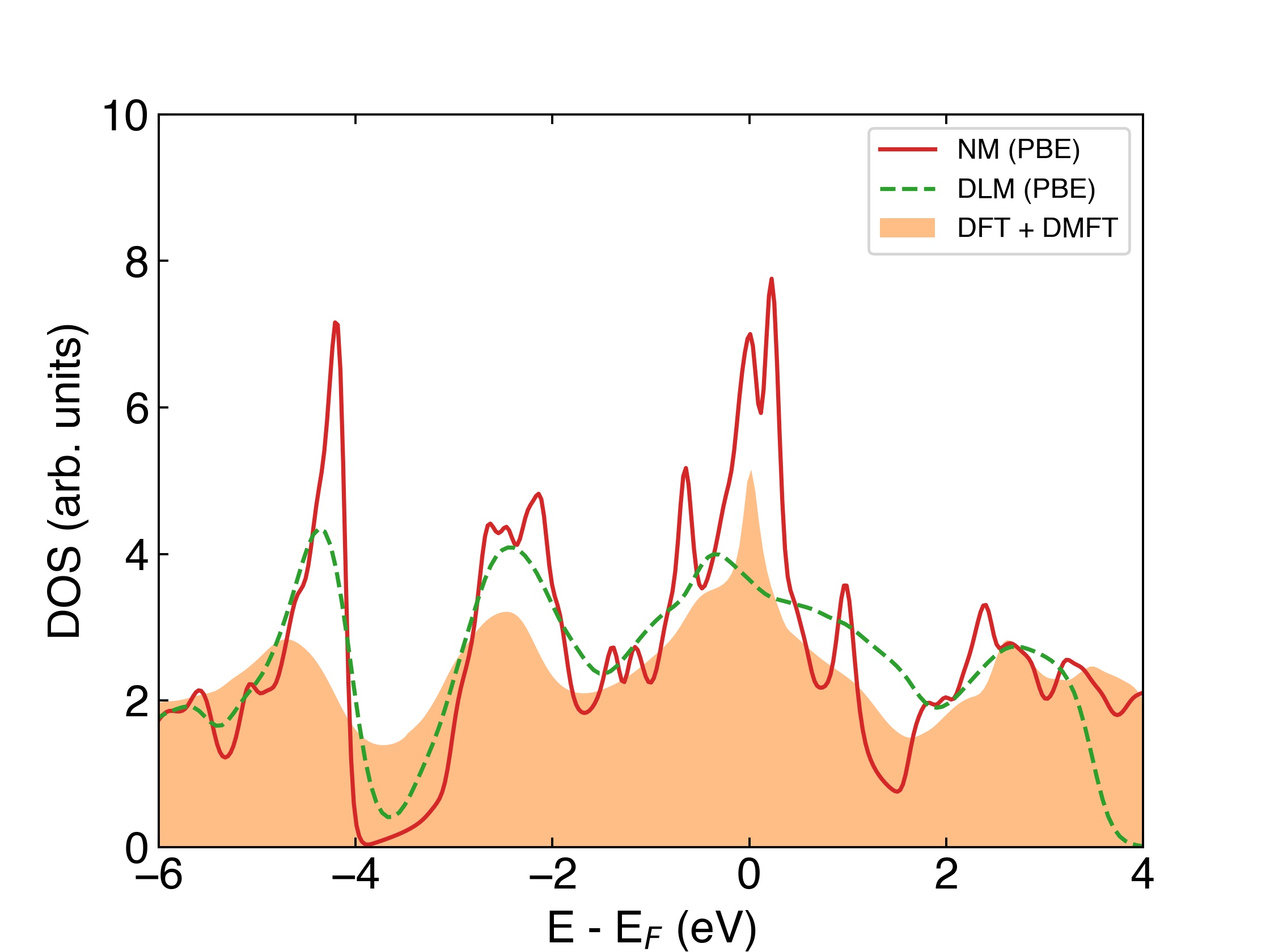}
\caption{Density of states of Mn$_2$GaC obtained within nonmagnetic PBE (solid red line), disordered local moments within the PBE approximation (dashed green line), and the DFT+DMFT total spectral function (solid orange field). The energy scale is relative to the Fermi energy, $E_F$.}
\label{fig:alldos}
\end{figure}

In Fig.\ \ref{fig:alldos} we compare the density of states (DOS) obtained within DFT with the {\bf k}-integrated spectral function from DFT+DMFT.
Besides DLM, we also include nonmagnetic DFT (PBE) calculations.
We note that in the DFT-DLM results the main characteristic features of the nonmagnetic DOS are still present, and the electronic structure of all local environments add up to a very similar DOS as in the nonmagnetic DFT description.
This is in line with the fact that the DFT+DMFT calculations retain the large quasiparticle peak due to the Mn $3d$ states near the Fermi energy that is seen in the DFT results, although it is significantly renormalized due to correlation effects.

\section{Discussion}

The electronic structure and magnetic properties of MAX-phases have received much attention in condensed matter physics during the last decade. Nonetheless, the nature of their magnetic interactions, e.g., the origin of magnetic moments of the prototypical magnetic MAX-phase compound Mn$_2$GaC has remained controversial. In our study, we compute the electronic structure and magnetic properties of the PM, FM, and AFM phases of Mn$_2$GaC using the DFT+DMFT and DFT-DLM electronic structure methods. While DFT-DLM gives a static configurational treatment of the PM state, by using DFT+DMFT it becomes possible to treat dynamical quantum fluctuation effects at finite temperatures and to study local moments formation in Mn$_2$GaC.

%
For the PM phase using DFT+DMFT we find a correlated metal with a Fermi-liquid-like behavior of the self-energy exhibiting a large orbital-selective quasiparticle damping (for the Mn $t_2$ and $e_g^\sigma$ states $\mathrm{Im}[\Sigma(0^+)]\sim0.34$ and $\sim$0.14 eV in the PM state at 290~K, respectively) and remarkable orbital-selective renormalization of the Mn $3d$ bands. It leads to a sufficiently strong orbital-selective incoherence of the spectral weight of the Mn $3d$ states near the Fermi level, implying the importance of electronic correlation effects. 
Our DFT+DMFT results reveal a complex magnetic behavior of Mn$_2$GaC with strongly competing FM and AFM states at low temperatures. This suggests a high sensitivity of the magnetic state of Mn$_2$GaC to fine details of its crystal structure, lattice volume, pressure, etc., in agreement with recent experiments. Moreover, this suggests the possible importance of structural optimization of the atomic positions and unit-cell shape of Mn$_2$GaC within DFT+DMFT, which remains a great challenge for the future \cite{PhysRevLett.112.146401,PhysRevLett.126.106001,PhysRevB.94.195146,Haule_2018,PhysRevLett.119.067004}. While the ordered magnetic moments obtained by DFT+DMFT for the FM and AFM states ($\sim$1.7$\mu_B$) are compatible with those found with spin-polarized DFT, the instantaneous magnetic moment is sufficiently higher ($\sim$2.5$\mu_B$), implying the presence of significant (local) spin fluctuations. Our analysis of the local spin susceptibility suggests the proximity of the Mn $3d$ moments to localization, consistent with a large quasiparticle damping of the Mn $t_2$ orbitals in the PM state. At the same time, the quasiparticle band renormalization of the Mn $3d$ states in the long-range magnetically ordered FM and AFM phases is found to be relatively weak, about 1.5.

We observe that the magnetic properties of Mn$_2$GaC depend very sensitively on the choice of the Hund's exchange coupling, with a typical itinerant magnetism and coherent quasiparticle behavior for $J=0.5$~eV and robust local moments behavior for $J=0.95$~eV. Moreover, DFT+DMFT calculations with $J=0.5$~eV give a sharp suppression of the magnetic ordering temperature to $\sim$116~K, resulting in a weak ordered magnetic moment of $\sim$0.12$\mu_B$ per Mn ion (at $\sim$116~K), in contradiction with experiment.
Most importantly, our calculations show that the magnetic properties of Mn$_2$GaC are dictated by its proximity to the regime of formation of local magnetic moments, in which localization is driven by the Hund's exchange coupling $J$. We observe that the quantum dynamics of the system is driven by the Hund's exchange instead of the Coulomb repulsion, suggesting that Mn$_2$GaC is a representative of Hund’s metals. This results in a remarkable orbital-dependent incoherence of the spectral weight of PM Mn$_2$GaC, which is different to the DFT-based DLM results.

Our DFT+DMFT calculations suggest the robustness of local magnetic moments of the Mn ions in this compound. 
This explains why the previous DFT calculations give reliable results for the long-range ordered magnetic state of Ma$_2$GaC.
%
%
The robust local moment behavior seems to validate the use of a static configurational DFT-DLM treatment of the PM state of Mn$_2$GaC, while the magnitude of the localized magnetic moments is found to be affected by temperature-induced excitations.
Moreover, the DFT-based description of the high-temperature PM state of Mn$_2$GaC in the framework of the static mean-field DLM picture leads to a mean magnetic moment of $\sim$1.7$\mu_B$, compatible with the fluctuating local magnetic moments obtained by DFT+DMFT. 
At the same time, the DFT+DMFT calculations suggest the presence of significant (local) spin fluctuations in Mn$_2$GaC. This effect seems to be partly captured in our DFT-DLM calculations where a significant portion of the magnetic moments deviates from the mean value.

\section{Summary and Conclusions}

In conclusion, we performed a theoretical study of the electronic structure and magnetic properties of the PM, FM, and AFM states of the prototypical MAX-phase Mn$_2$GaC using the DFT+DMFT and DFT-DLM methods. Our DFT+DMFT results show robust local-moment behavior and orbital-selective incoherence of the spectral properties of Mn$_2$GaC, which imply the importance of orbital-dependent localization of the Mn $3d$ states.
This suggests the crucial importance of (orbital-dependent) correlation effects for understanding the electronic and magnetic properties of Mn$_2$GaC. 

Our calculations reveal a complex magnetic behavior of Mn$_2$GaC with strongly competing FM and AFM states at low temperatures. This suggests a high sensitivity of the magnetic state of Mn$_2$GaC to fine details of its crystal structure, lattice volume, pressure, etc., in agreement with recent experiments.
This is in agreement with the results of the previous DFT calculations of the long-range magnetically ordered states of Mn$_2$GaC, while the robust local moment behavior seems to validate the use of a static configurational DFT-DLM treatment of the PM state of Mn$_2$GaC.

Most importantly, our calculations show that the magnetic properties of Mn$_2$GaC are dictated by its proximity to the regime of formation of local magnetic moments, in which localization is driven by the Hund's exchange coupling $J$. We observe that the quantum dynamics of the system is driven by the Hund's exchange instead of the Coulomb repulsion, suggesting that Mn$_2$GaC is a representative of Hund’s metals.
Our results may have important implications for the theoretical and experimental understanding of the magnetic properties of MAX phases. We believe that this topic deserves further detailed theoretical and experimental considerations.

\begin{acknowledgments}
Support from the Knut and Alice Wallenberg Foundation (Wallenberg Scholar Grant No.\ KAW-2018.0194), the Swedish Government Strategic Research Areas in Materials Science on Functional Materials at Linköping University (Faculty Grant SFO-Mat-LiU No.\ 2009 00971), the Swedish e-Science Research Centre (SeRC), the Swedish Research Council (VR) grant No.\ 2019-05600 and Swedish Foundation for Strategic Research (SSF) Project No.\ EM16-0004 is gratefully acknowledged. Theoretical analysis of the DFT+DMFT calculations was supported by the Russian Science Foundation (Project No.\ 18-12-00492). Analysis of DFT results was supported by the state assignment of Minobrnauki of Russia (theme ``Electron'' No. AAAA-A18-118020190098-5). The computations were carried out at resources provided by the Swedish National Infrastructure for Computing (SNIC) partially funded by the Swedish Research Council through grant agreement no.\ 2016-07213 and at supercomputer cluster at NUST ''MISIS''. 
\end{acknowledgments}

\bibliographystyle{apsrev4-2}
\bibliography{mn2gac}
\end{document}